\documentclass{iopconfser}
\usepackage{natbib}
\usepackage{graphicx}
\usepackage{xcolor}
\usepackage{colortbl}

\begin{document}

\title{Quantum Cosmology and the Age of the Universe}

\author{\'Alvaro Mozota Frauca}

\affil{Department of Condensed Matter Physics. Universitat de Barcelona, Martí i Franquès, 1. 08028 Barcelona (Spain) \\
Department of Mathematics. Universitat Polit\`ecnica de Catalunya, Pau Gargallo, 14. 08028 Barcelona (Spain)}

\email{alvaro.mozota@upc.edu}

\begin{abstract}
In this article I study how the problem of time of canonical approaches to quantum gravity affects the simple minisuperspace models used in quantum cosmology. I follow some authors who have argued that this issue makes the quantization of general relativity problematic to conclude that the same applies in the case of quantum cosmology. In particular, I argue that temporal structures are lost in quantization and that this is a problem, as they encode part of the empirical content of classical cosmology, such as the age of the universe.
\end{abstract}

\section{Introduction}

Quantum cosmology applies the techniques developed by the quantum gravity community to simple cosmological models to obtain quantum theories that aim to apply to the whole universe at cosmological scales, especially at very early times, when quantum effects are believed to be relevant. Works in quantum cosmology suggest interesting ideas such as time being discrete or replacing the Big Bang with a Big Bounce.\footnote{See \citep{Calcagni2017} and the references therein for an overview of the field.} However, by adopting the formal structures of full quantum gravity, quantum cosmology also suffers from its conceptual shortcomings. In this article I follow the authors\footnote{In particular see \citep{Kuchar1993,Gryb2016,MozotaFrauca2023,mozota_frauca_problem_2024}.} that have argued that the problem of time is a serious issue that jeopardizes the justification and interpretation of canonical approaches to quantum gravity to argue that the same applies to the case of canonical quantum cosmology. In particular, I argue that a signal of this is that the facts about the age of the universe, which is an important empirical fact in the classical models, go missing in the quantum ones.

The issues discussed in this article are of interest to both physics and philosophy of physics, as they concern foundational issues in approaches to quantum gravity.\footnote{See the volumes \citep{Huggett2020,Wuthrich2021} for recent pieces discussing some topics in the foundations of quantum gravity.} This discussion is certainly worth considering from the perspective of these approaches and it also provides a technically more simple context in which the conceptual discussion of the problem of time can be carried out without getting misled by the technical complications involved with dealing with full general relativity.

The structure of this article is the following. I start in section \ref{sect_problem} by introducing the problem of time and by giving an overview of the argument that it is a serious issue in the quantization of certain models. Then, in section \ref{sect_cosmo} I introduce classical minisuperspace models and argue that their quantization suffers from a problem of time. Finally, in section \ref{sect_interpretations} I study the ways in which the quantum cosmology community has interpreted the quantum cosmological models and I analyze how the problem of time makes them unappealing and unable to recover the temporal structures of the classical theory. I also comment on the semiclassical approach, which is able to postulate some temporal structures for certain states, but which doesn't give an interpretation for the full theory.

\section{The problem of time}\label{sect_problem}

In this section I give a brief overview of the problem of time that appears in the canonical quantization of any reparametrization invariant system. The discussion is divided into two parts. First, I discuss the problem in general and then I study how it affects the quantization of three different models for one system and argue it represents a serious issue for models like general relativity.

\subsection{The problem}

In classical theories we find that time plays two fundamental roles.\footnote{I am leaving out of the discussion whether time plays further roles such as defining an arrow of time, which are not necessary for my discussion in this article.} First, it defines a chrono-ordinal structure which allows describing what happens in the world as a sequence of instants or events. For instance, we describe sequences of facts such as me kicking a ball, its traveling through space, and its falling at a certain distance. Second, time also plays a metric role: it defines durations. In our example we not only can describe a sequence of positions that the ball takes, but we can also claim that this process takes some amount of time. 

In relativistic theories, both special and general relativistic, we find that temporal structures play the same two roles, even if in a more complex way. The causal structures of spacetime define a partial order relation between events, which means that we can still describe physical processes as sequences of events. Similarly, along any time-like curve in spacetime there is a well-defined proper time, so durations can also be defined. In other words, the example of the ball being kicked can also be represented in a relativistic setting as a ball traveling through spacetime between the moment it is kicked and the moment it touches the floor and one can also compute the time this process lasts (as experienced by the ball).

In quantum theories, relativistic or not, we also find the same two structures, and they are a fundamental part of the way we understand these theories. The problem of time is precisely the disappearance of the temporal structures of a classical theory once one applies canonical quantization techniques to certain theories.\footnote{See the two classical reviews on the topic \citep{Isham1993,Kuchar1992} and the more recent one \citep{Anderson2017}. Some authors like Anderson distinguish different but related problems of time. By `problem of time' in this article I am mostly referring to what Anderson calls the `frozen formalism problem'.} This disappearance is what some authors\footnote{See \citep{Kuchar1993,Gryb2016,MozotaFrauca2023,mozota_frauca_problem_2024}.} have argued that makes models affected by the problem of time unappealing and to raise the worry that the quantization has failed to give a meaningful theory.

More precisely, the problem of time arises in the quantization of any temporal reparametrization invariant model, that is, any model in which the temporal parameter or coordinate can be freely chosen. This is of course the case of general relativity, where one has an explicit freedom to choose any arbitrary coordinate system to describe spacetime. In formal terms, these models describe how a series of configuration variables $q_1, q_2...$ evolve with respect to an arbitrary parameter $\tau$. Given a solution of the equations of motion $q_1(\tau), q_2(\tau)...$ one can obtain an equivalent one by transforming the temporal parameter, that is, by replacing $\tau$ with $f(\tau)$, where $f$ is any monotonic function. 

When one follows the canonical quantization procedure, the symmetry of this kind of model implies that quantum states $\psi(q_1,q_2...;\tau)$ need to satisfy a constraint equation of the form:
\begin{equation}\label{eqn_constraint}
\hat{\mathcal{H}}\psi(q_1,q_2...;\tau)=0 \, ,
\end{equation} 
where $\hat{\mathcal{H}}$ is an operator related to the Hamiltonian of the system.\footnote{$\mathcal{H}$ is a phase space function known as the Hamiltonian constraint, and the Hamiltonian is proportional to it, up to the addition of other constraints. See \citep{MozotaFrauca2023} for a more technical description of the formalism.} This equation implies that the Schr\"odinger equation for the system becomes trivial:
\begin{equation}
\frac{\partial}{\partial \tau} \psi(q_1,q_2...; \tau)=\hat{\mathcal{H}}\psi(q_1,q_2...;\tau)=0 \, .
\end{equation} 
In other words, what one obtains by applying canonical quantization methods to reparametrization invariant systems is quantum states that are independent of the time parameter of the evolution. For this reason one finds claims such that quantum states are frozen or that time has disappeared from the formalism. 

This constitutes the problem of time for reparametrization invariant systems. However, the severity of this problem depends on the exact form of these systems, as has been argued in \citep{MozotaFrauca2023}. That is, while in some theories the loss of the dependence on the time parameter represents a serious issue, in others it does not, as long as the empirical content encoded in this dependence is also encoded in some other variables of the model. This is the case of deparametrizable models, which have a redundancy that allows claiming that the problem of time is not fatal. Unfortunately, models like general relativity or cosmological models are not of this kind, and the problem becomes much more serious. In the following subsection I discuss a concrete example that will make this argument clearer.

\subsection{One system, three models}

\begin{figure}
\begin{center}
\includegraphics[scale=0.6]{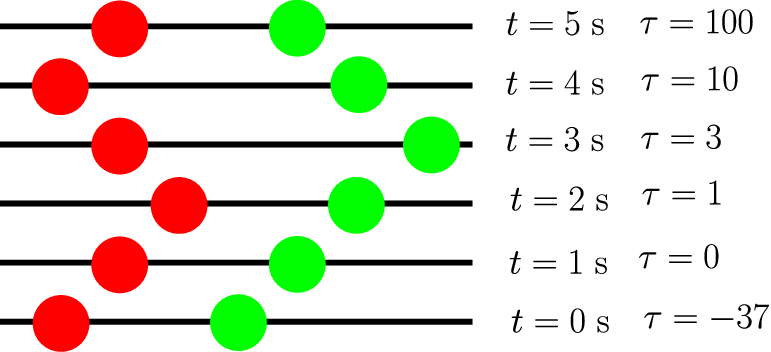}
\end{center}
\caption{\label{fig_particles} A system of 2 particles in one spatial dimension evolving in time. Each horizontal line represents the state of the system at a time, i.e., the position of the two particles on the line. These instants are labeled by the Newtonian time $t$ and an arbitrary parameter $\tau$. The models discussed in this section will differ in whether they use $t$ or $\tau$.}
\end{figure}

Consider a Newtonian system of two particles in one spatial dimension (figure \ref{fig_particles}) subject to two different potentials $V_1(x_1)$ and $V_2(x_2)$. By assumption this is a model in a Newtonian spacetime satisfying Newton's equations. Now, this system can be described by three different actions, which correspond to three different ways of representing time in Newtonian physics.\footnote{See \citep{mozota_frauca_time_2024} for a discussion of the role that time variables play in our physical theories.} Whether we find a problem of time when quantizing these actions or not will depend on these different ways of representing time. I will now discuss these models and their quantizations. Table \ref{table_models} summarizes the relevant facts.

\begin{table}[]
\resizebox{\textwidth}{!}{
\begin{tabular}{|l|l|l|l|l|l|}
\hline
\rowcolor[HTML]{C0C0C0} 
\multicolumn{1}{|c|}{\cellcolor[HTML]{C0C0C0}Model} & \multicolumn{1}{c|}{\cellcolor[HTML]{C0C0C0}\begin{tabular}[c]{@{}c@{}}Reparam. \\ inv?\end{tabular}} & \multicolumn{1}{c|}{\cellcolor[HTML]{C0C0C0}\begin{tabular}[c]{@{}c@{}}Time\\  (order)\end{tabular}} & \multicolumn{1}{c|}{\cellcolor[HTML]{C0C0C0}\begin{tabular}[c]{@{}c@{}}Time\\  (metric)\end{tabular}} & \multicolumn{1}{c|}{\cellcolor[HTML]{C0C0C0}Quantization} & \multicolumn{1}{c|}{\cellcolor[HTML]{C0C0C0}Problem?} \\ \hline
\begin{tabular}[c]{@{}l@{}}$x_1(t)$\\ $x_2(t)$\end{tabular}                                             & No                                                                                                    & $t$                                                                                                    & $t$                                                                                                     & \begin{tabular}[c]{@{}l@{}}$i\hbar \frac{\partial \psi}{\partial t} = \hat{H}\psi$\\ $\psi(x_1,x_2;t)$\end{tabular}         & No                                                    \\ \hline
\begin{tabular}[c]{@{}l@{}}$x_1(\tau)$\\ $x_2(\tau)$ \\$t(\tau)$\end{tabular}                                                            & Yes - Deparam.                                                                                                   & $\tau$ (and $t$)                                                                                       & $t$                                                                                                     &\begin{tabular}[c]{@{}l@{}}$i\hbar \frac{\partial \psi}{\partial \tau} = \hat{\mathcal{H}}\psi=0$\\ $\psi(x_1,x_2,t)$\end{tabular}         & No                                                    \\ \hline
\begin{tabular}[c]{@{}l@{}}$x_1(\tau)$\\ $x_2(\tau)$\end{tabular}                                                          & Yes  - Non-Deparam.                                                                                                 & $\tau$                                                                                               & \begin{tabular}[c]{@{}l@{}}Not explicit\\ \\ $dt=\sqrt{\frac{m_1\dot{x}_1^2+m_2 \dot{x}_2^2 }{2E-2V_1(x_1)-2V_2(x_2)}}d\tau$\end{tabular}                                                     & \begin{tabular}[c]{@{}l@{}}$i\hbar \frac{\partial \psi}{\partial \tau} = \hat{\mathcal{H}}\psi=0$\\ $\psi(x_1,x_2)$\end{tabular}          & Yes                                                   \\ \hline
\end{tabular}}
\caption{\label{table_models}Summary of the three models discussed for the Newtonian system of two particles and its quantization.}
\end{table}

First, we have the standard Lagrangian for the system:
\begin{equation}
S_{1}[x_1,x_2]=\int dt \left(\frac{m_1}{2}\dot{x}_1^2+\frac{m_2}{2}\dot{x}_2^2-V_1(x_1)-V_2(x_2)\right) \, .
\end{equation}
In this action the time parameter is Newtonian time $t$ and imposing that physical trajectories $x_1(t),x_2(t)$ minimize it leads to Newton's equations of motion.

Second, we have a reparametrization invariant version of the same model:
\begin{equation}
S_2[x_1,x_2,t]=\int d\tau \left(\frac{m_1}{2}\frac{\dot{x}_1^2}{\dot{t}}+\frac{m_2}{2}\dot\frac{{x}_2^2}{\dot{t}}-\dot{t}V_1(x_1)-\dot{t}V_2(x_2)\right) \, .
\end{equation}
In this action the time parameter is not Newtonian time $t$, but an arbitrary parameter $\tau$, and the dot (e.g. $\dot{x}_1$) represents a derivative with respect to $\tau$. Newtonian time appears as a configuration variable, and the model exhibits an explicit reparametrization invariance. That is, given a solution of its equations of motion $x_1(\tau),x_2(\tau),t(\tau)$, one can find an equivalent one by replacing $\tau$ with $f(\tau)$. In this model we have a redundancy, as both $\tau$ and $t$ encode facts about time. In particular, while $\tau$ is the parameter of evolution and one can readily read the chrono-ordinal structure of time from it, $t$ also encodes this information. The metric aspect of time is only encoded in $t$, and one can read it by looking at the function $t(\tau)$.  One can even go one step further and invert this relationship to obtain the function $\tau(t)$, and by substituting it in $x_1(\tau),x_2(\tau)$, recover the solutions in Newtonian time $x_1(t),x_2(t)$. This process of eliminating the arbitrary parameter $\tau$ by inverting $t(\tau)$ is called deparametrization, and this kind of model, deparametrizable.

Finally, we have a third model, which is reparametrization invariant but not deparametrizable:
\begin{equation}
S_3[x_1,x_2]=\int d\tau  \sqrt{ 2\left( m_1\dot{x}_1^2+m_2 \dot{x}_2^2 \right)\left(E-V_1(x_1)-V_2(x_2)\right)} \, ,
\end{equation} 
where $E$ is a constant representing the energy of the system. In this case we do not have a redundancy, as $t$ has completely disappeared from the picture, although the equivalence with the original model is preserved. The reason for this is that the solutions of the equations of motion of this model, $x_1(\tau), x_2(\tau)$ are equivalent to the Newtonian trajectories. The difference is that now the trajectories are parametrized in an arbitrary way, $\tau$ is not Newtonian time, and we can change the parametrization and still obtain an equivalent solution. $\tau$ encodes the chrono-ordinal structure of time, while the metric one seems to be missing, as $t$ is not present in this kind of model. For this reason, this kind of model is called non-deparametrizable.

However, metric Newtonian time can be recovered in this model. The way of doing so is by applying the following relation between coordinate time $\tau$ and Newtonian time $t$:
\begin{equation}
dt= \sqrt{\frac{m_1\dot{x}_1^2+m_2 \dot{x}_2^2 }{2E-2V_1(x_1)-2V_2(x_2)}} d\tau \, .
\end{equation} 
This relationship is similar to the way proper time along a spacetime trajectory is defined in a relativistic spacetime:
\begin{equation}
dt=\sqrt{g_{\mu\nu}\dot{x}^{\mu}\dot{x}^{\nu}}d\tau
\end{equation} 
In this sense, we can already see how this model is the one that represents the spacetime structure of Newtonian spacetime in the closest way to how spacetime structure is represented in general relativistic physics.

Let us discuss the quantization of each of these models.

For the first model, canonical quantization techniques work fine and deliver the standard quantum theory one would be expecting, i.e., quantum states of the form $\psi(x_1,x_2;t)$ which evolve with respect a Schr\"odinger equation with Hamiltonian $\hat{H}=\frac{\hat{p}_1^2}{2m_1}+\frac{\hat{p}_2^2}{2m_2}+V_1(\hat{x}_1)+V_2(\hat{x}_2)$. As there is no reparametrization invariance there is no problem with this kind of model.

For the second model, we find a problem of time. Naively, we would expect quantum states of the form $\psi(x_1,x_2,t;\tau)$, but the symmetry of the model makes it the case that states become independent of $\tau$. However, states are still of the form $\psi(x_1,x_2,t)$ and one can argue that they are still time-dependent, as there is some dependence on $t$. That is, given the redundancy that we had in the classical model, the loss of $\tau$ is not fatal, as we still have $t$ to play the role of time. Furthermore, the constraint equation (\ref{eqn_constraint}) turns out to be equivalent to the Schr\"odinger equation of the previous model. In this sense, the quantization of our second model can be argued to be equivalent to the quantization of the first one.

For our third model, the situation is not so positive. Quantization leads to a problem of time in which we were expecting states of the form $\psi(x_1,x_2;\tau)$ but what we obtain is that these states are independent of $\tau$, that is, $\psi(x_1,x_2)$. It is straightforward to see that this quantization is formally nonequivalent to the ones of the other two models, as there is no straightforward way of claiming that $\psi(x_1,x_2)$ can be equivalent to $\psi(x_1,x_2;t)$ or $\psi(x_1,x_2,t)$. Similarly, the constraint equation (\ref{eqn_constraint}) in this case is not equivalent to a time-dependent Schr\"odinger equation, but it rather imposes that states $\psi(x_1,x_2)$ are energy eigenstates. In this sense, the quantization has produced something different in this case.

Can we still claim that the outcome of the quantization is a meaningful quantum theory? \citep{MozotaFrauca2023} argues against this possibility. Clearly, the result of the other two quantizations is the standard quantization, for which we have good reasons to believe, a way of connecting it with measurements, and ways of connecting it with the classical theory we started with. For the `timeless' quantization of the third model, we lack all of these, and the lack of temporal dependence makes it the case that an important part of the empirical content of the classical theory seems to go missing in the quantum one. In this sense, I believe that there are good reasons for believing that this third quantization has failed to give a satisfactory quantum version of the original system we started with. I will expand these arguments and study the way they apply to the cosmological case in section \ref{sect_interpretations}. Before this, notice that while some authors argue that the problem of time arises in general relativity or cosmology because we are quantizing spacetime, in the model discussed in this section this wasn't the case, and we still had a problem of time. The problem of time appears for certain representations of temporal structures, independently of whether there is a sense in which we are quantizing spacetime.

\section{Cosmology}\label{sect_cosmo}

The cosmological models I am analyzing in this article are the simple minisuperspace models and their quantization, which is a wide class of models discussed in the quantum cosmology literature. I start by discussing the classical models and their temporal structures before moving on to their quantization. For a recent review of the field, I refer the reader to \citep{Calcagni2017}.

\subsection{Classical cosmology}

Contemporary cosmology describes space and time as a relativistic spacetime making use of the tools of general relativity. However, given that our universe is highly symmetric, especially so for the very early universe, many cosmological models are defined on a class of symmetry-reduced models. That is, as it is a good approximation to assume that the universe can be described as homogeneous and isotropic at cosmological scales, cosmologists build models with these properties, which is a considerable simplification. These spacetime models are the FLRW spacetimes, and their line element can be written as:
\begin{equation}\label{eqn_FLRW}
ds^2=-N(\tau)^2d\tau^2+a^2(\tau)(dx^2+dy^2+dz^2) \, .
\end{equation}
In this form, it is explicit that there is only one geometrical degree of freedom, the scale factor $a$, and that it has only temporal dependence. Similarly, the other degrees of freedom that there may be in the model are assumed to be homogeneous and isotropic when described in these coordinates. The dynamics of general relativity are therefore simplified from a set of 10 partial differential equations to just two ordinary differential equations, which are known as Friedmann equations. These equations describe how the universe expands or contracts depending on the matter and energy content of the universe and are widely used by cosmologists.

As a relativistic spacetime, in any FLRW spacetime we find the two types of temporal structures discussed in the previous section. That is, the line element (\ref{eqn_FLRW}) defines a causal structure that establishes a partial order relation between events and also the proper time that any observer, following any trajectory in spacetime would experience. However, the symmetries of the model make it convenient to choose a particular time coordinate for describing it. This coordinate is known as cosmic time $t$, and it is defined as the proper time that an observer at rest with respect to the frame in which space is homogeneous and isotropic would experience. It is the time variable that allows making claims like that the age of the universe is 13.8 billion years. This sort of claim is of course part of the empirical content of our cosmological models, and it is usually treated as such. That is, a model which predicted that an observer at rest with respect to the foliation which makes spacetime homogeneous and isotropic would have experienced a different duration would be wrong, and it would lead to wrong predictions regarding the abundance of certain elements or the population of certain bodies like stars or black holes.

Instead of writing the line element (\ref{eqn_FLRW}) directly in terms of cosmic time, I have chosen to write it in terms of an arbitrary time coordinate $\tau$ and the lapse function $N(\tau)$, which relates both time coordinates $dt=N(\tau)d\tau$. This makes explicit the temporal reparametrization invariance of the model, which is non-deparametrizable. That is, this kind of model describes the evolution of a series of variables $a,N,\rho...$ evolving with respect to an arbitrary parameter $\tau$, and the metric time $t$ does not appear as a configuration space variable. In this sense, cosmological models are similar to the third model in the previous subsection, and their quantization will be similar.

%
%
%
%
%
%

\subsection{Canonical quantum cosmology}

Canonical quantum cosmology proceeds by applying canonical quantization methods to FLRW models. In this way, one obtains wavefunctions on the configuration space of these models. That is, one gets wavefunctions of the form $\psi(a,\phi^{\alpha})$, where $\phi^{\alpha}$ represents the matter degrees of freedom of the model. The space where these wavefunctions are defined is known as minisuperspace, as it is the symmetry-reduced version of superspace, the configuration space of general relativity. For this reason, this kind of model is known as minisuperspace model.

There are different types of minisuperspace models depending on technical details of which variables to use as a basis for quantization or whether one uses metric variables like $a$ or some other classically equivalent formulation, such as in the holonomy-based formulations of minisuperspace models.\footnote{This is the cosmological version of loop quantum gravity, known as quantum cosmology. See \citep{Agullo2017,Ashtekar2009}.} These choices lead to different quantizations and have been interpreted to have different consequences. For instance, whether there is a Big Bang singularity in the quantum has been argued to depend on this sort of choice. Similarly, some other authors have argued that whether in the resulting quantum theory there is a discrete or continuous time also depends on the exact quantization carried out.\footnote{See \citep{Bojowald2004,Ashtekar2006}.} In this sense, there are many interesting potential consequences in the technical choices involved in the definition of the minisuperspace and Hilbert space associated. However, in this work I am not interested in studying those, as my focus is on the problem of time of these approaches, which is independent of these issues. That is, independently of the choices of variables involved in the quantization, at the end of the day one faces a problem of time.

Just as I explained in section \ref{sect_problem}, as cosmological models are reparametrization invariant, instead of having wavefunctions $\psi(a,\phi^{\alpha};\tau)$ which depend on the evolution parameter $\tau$, the quantization procedure produces $\tau$-independent wavefunctions $\psi(a,\phi^{\alpha})$. These states have to satisfy a certain constraint equation\footnote{In the case of full general relativity, this constraint equation is the Wheeler-deWitt equation, and some authors use the same name to refer to the constraint equation in the case of cosmology.} but there is no (non-trivial) dynamical equation describing how they evolve in $\tau$. This is just the problem of time for minisuperspace models.

Notice that as cosmological models are non-deparametrizable, that is, as there is no time variable in their configuration space, the situation is just analogous to model 3 in section \ref{sect_problem}. That is, as there is no redundancy, there is no straightforward way of reinterpreting the wavefunctions $\psi(a,\phi^{\alpha})$ as secretly representing a time-evolution as we did for model 2. In this sense, the problem of time is serious, as facts about time, including the prediction of the age of the universe, seem to go missing. In the next section I analyze the ways these timeless states have been proposed to be interpreted in the quantum cosmology literature and their shortcomings.

\section{Interpretations of quantum cosmology and their shortcomings}\label{sect_interpretations}

In the quantum cosmology literature we find two main ways of interpreting the formalism: the relational interpretation and probabilistic interpretations. I will now discuss them and how the problem of time affects them. I also briefly discuss the semiclassical interpretation, which is the one that gets closer to recovering facts about time, but I argue that it gets short of constituting an interpretation of the full quantum formalism.

\subsection{Relational interpretation}

The most extended interpretation of canonical quantum cosmology is the relational interpretation.\footnote{Some examples of the application of this strategy are \citep{Bojowald2004,Ashtekar2006,Oriti2017,Gielen2022,Gielen2022a}.} According to this view, time variables in physical theories just encode relations between physical and measurable variables and time coordinates are meaningless. From this perspective, the loss of time coordinates is not fatal according to this view, as long as one can still define the evolution of some physical variables with respect to others. In this sense, timeless quantum states are argued to encode relational evolution.

A common strategy is to pick one of the variables in the formalism to act as a `clock' variable, that is, to be the variable with respect to which the rest of variable and the quantum state evolve. For instance, in the models containing just a scalar field $\phi$, the states $\psi(a,\phi)$ are commonly reinterpreted to be $\psi(a;\phi)$, that is, a quantum state for $a$ evolving in `time' $\phi$. In this way, the constraint equation is interpreted as a dynamical equation describing the way $\psi(a)$ evolves. In this sense, one is able to replicate the structures of standard quantum mechanics with $\phi$ playing the role that usually plays $t$ in quantum mechanics.

The motivation for adopting this strategy is that it works for deparametrizable models like the second model for the two-particle system in section \ref{sect_problem}. As I mentioned then, states like $\psi(x_1,x_2,t)$, even if they don't depend on the evolution parameter $\tau$, aren't timeless after all, as they carry a dependence on the configuration space variable $t$. In this way, the states are interpreted as states for $x_1$ and $x_2$ evolving in $t$.  

There are several problematic issues with this sort of idea. From my point of view, the most pressing is that even if one wants to be a relationalist, time variables and variables representing possible `clocks' represent fundamentally different structures and that proponents of the relational interpretation tend to conflate them. That is, the reason why this strategy works for deparametrizable models is that there is a variable in their configuration space that represents time, while for non-deparametrizable models like cosmological models there isn't. By defining evolution with respect to, say, $\phi$, what the relational strategy is doing is converting this `clock' variable into a time variable, which is conceptually mistaken. 

For instance, consider the example of the system of two particles described in section \ref{sect_problem}. Imagine that the potentials $V_1$ and $V_2$ are such that for the energies of the particles their motions are bounded in the intervals $[a_1,b_1]$ and $[a_2,b_2]$ for particles 1 and 2, respectively. Relationalists are right in that during the time in which particle 1 travels from $a_1$ to $b_1$ it may make sense to use its position as a clock for particle 2. That is, we may ask what will the position of particle 2 be when particle 1 is at a position $x_1$ and even express this as a function $x_2(x_1)$. Similarly, we could build a quantum theory in which we had a wavefunction for particle 2 evolving with respect to time as expressed as the position of particle one, i.e., $\psi(x_2;x_1)$.

These relationalist ideas work fine for just a limited amount of time, the time it takes particle 1 to travel from $a_1$ to $b_1$ once. But when we consider longer periods of time, particle 1 ceases to work as a good clock, in the sense that specifying a value of its position no longer works for identifying a moment in the evolution. In this sense we cannot simply ask what the position of particle 2 will be when particle 1 is at position $x_1$, we also need to specify which of the times particle 1 is at that position we are referring to. That is, particle 2 will be, for generic conditions, at different positions each time particle 1 reaches position $x_1$. In the quantum case, we expect that the wavefunction for particle 2 will be different each time particle 1 is at a certain position. This shows how the relationalist idea of taking one degree of freedom as a clock or time variable is of limited application.

Similarly, one could worry that it is not only the chrono-ordinal structure of time that is misrepresented in the relational interpretation of quantum cosmology or models like the non-deparametrizable version of the two-particle system. The metric aspect of time of our classical models seems to just disappear in these timeless states. Time in our classical models allowed making the claim that particle 1 traveled from $a_1$ to $b_1$ in a time $T$ or that the expansion of the universe from the CMB release to our days took 13,8 billion years. This sort of claim goes missing in the relational interpretation of states like $\psi(x_1,x_2)$ and $\psi(a,\phi)$. That is, if we take, say, $\phi$ to be the clock variable we can claim that the state evolves from $\psi_1(a)$ to $\psi_2(a)$ in the time $\phi$ goes from a value $\phi_1$ to another one $\phi_2$, but the relational strategy doesn't provide us with any way of translating this into an actual duration.

Another way of arguing that there is something conceptually mistaken with the relational interpretation is by observing that one is forcing a temporal interpration into some variable in the formalism that didn't have that role in the classical model. Notice that in my discussion above $x_1$ and $x_2$ were treated very differently. While $x_1$ was considered to be evolving classically from $a_1$ to $b_1$, for the second particle we claimed to have an evolving quantum state $\psi(x_2)$. While the second particle is described as having a quantum dynamics, the first one is interpreted to be as classical and identifying instants of time. This is in contrast with the way in the classical theory both particles are treated equally. Both particles represent physical degrees of freedom that evolve on their own in the temporal structure of the model, which is something different from these degrees of freedom. Then, it seems conceptually wrong to interpret one of these degrees of freedom as a temporal structure or something very close to it when the actual temporal structure falls out of the quantum formalism.

Along the same lines, notice that one would naively expect to have phenomena like entanglement between the positions of both particles or between $a$ and $\phi$ in the case of cosmology. However, when we interpret one of them as a time variable a state that we would naively understand as being entangled now represents something like a time evolution. As one is effectively treating one of the variables as a classical variable the quantum phenomena like entanglement and superposition would be lost for them.

Finally, there is the issue of which variable to use as a `clock'. In the example of the two particles the position of any of them could be chosen, leading to two different theories. From my point of view, this shows how the relational interpretation is forcing a temporal interpretation into a variable that clearly didn't have anything to do with time. In the case of quantum cosmology exactly the same happens: choosing different variables as clocks leads to different theories, and this is problematic. For instance, \citep{Gielen2022,Gielen2022a} discuss a quantum cosmological model in which there are three possible choices of clock. Gielen and Men\'endez-Pidal show how depending on this choice the quantum theory one obtains is completely different: one of the options leads to a theory in which there isn't a Big Bang singularity at the very early universe but a Big Bounce after which the universe expands indefinitely, another one represents a universe in which the expansion reaches a maximum volume and then the universe recontracts, and the third option represents an expanding universe in which the Big Bang singularity is not avoided. Again, from my point of view this shows how in the relational interpretation one is forcing a temporal interpretation into variables that are not time. From a technical point of view one can build these theories, but there is no principled way of telling which one is the right one because none of them is, as any of them mistakenly takes a dynamical variable to play the roles that temporal variables take.

To sum up, in the relational interpretation of quantum cosmological states one tries to recover temporal structures of the original classical theory by reading them from the dynamical variables. I consider that this is a conceptual mistake, as one is reading a chrono-ordinal structure from some variables that didn't have anything to do with chrono-ordinal structure in the classical theory. Similarly, the metric aspect of time that in the classical theory allowed us to claim that the expansion of the universe takes 13.8 billion years just goes missing. 

\subsection{Probabilistic interpretations}

The most popular alternative to the relational interpretation of quantum cosmological states is the probabilistic interpretation. In this kind of approach, the timeless states are interpreted as encoding some sort of probability. These probabilities would allow us to make predictions and would constitute a meaningful quantum theory. However, from my point of view the way this would be so is unclear, and, in particular, the relationship with the original temporal structure is obscure to me, if there is any.

There are at least two ways the structures of quantum cosmology have been interpreted probabilistically. First, states $\psi(a,\phi)$ can directly be read as encoding the probability (density) of finding the universe with a size given by $a$ and the scalar field having the value $\phi$ just as a state $\psi(x)$ in quantum mechanics encodes the probability of finding a particle at some position $x$. That is, the amplitude square $|\psi|^2$ would encode this probability.\footnote{This interpretation of quantum cosmological models has been held in \citep{Rovelli2002,Colosi2003}.} Alternatively, one can use the inner product structure of the Hilbert space to define a mathematical object of the form $W(a,\phi;a'\phi')$, which is interpreted as the probability amplitude of finding the universe with properties $a,\phi$ given that it had properties $a',\phi'$.\footnote{This interpretation appears for instance in \citep{Colosi2003,Rovelli2015}.} I will now argue that both interpretations are problematic. 

Consider the first interpretation first. That is, consider that $|\psi(a,\phi)|^2$ represents some sort of probability density of finding the universe with a given size and with the scalar field taking some value. Leaving aside the technical worries that one may have about how this probability density is to be defined, from a conceptual point of view the claim isn't very illuminating when we try to connect it with our universe. As I have argued above, our world, at least in the classical regime we use to describe the universe at cosmological scales, is very well described by a series of degrees of freedom evolving in spacetime. Temporal structure is nowhere to be seen in the claim `$|\psi (a,\phi)|^2$ defines the probability of finding $a,\phi$'. It neither defines a chrono-ordinal structure to define evolution nor it allows recovering claims about duration. At most, it seems that it can be thought as distinguishing a `before' and an `after' the `measurement'. However, this is conceptually really unclear and there is no straightforward connection with our classical universe.

Let us turn to the second interpretation. In this second interpretation there is a more explicit temporal structure as there are two `moments' of time. However, the connection with the original temporal structure is still unclear. That is, even if we have two instants, this is far from having a continuum sequence as we had in the classical model. Moreover, the metric aspect of time is completely missing as one gives probabilities of finding a final state given that an initial one obtained, without making any reference to the amount of time that has elapsed from one moment to the other. In the case of cosmology we would like to be able to make claims like `given the initial conditions of the universe after the Big Bang, the most likely state after 13.8 billion years is that it has expanded $x$ times'. However, the proposal by the quantum cosmology community doesn't make any reference to how long it has been between both instants.

There are two attitudes that the quantum cosmology community takes towards interpreting these probabilities. In some works like \citep{Rovelli2015}, it seems that they consider that these probabilities are perfectly well-defined and understandable and they even define these probabilities for models in which the only degree of freedom is $a$. In some other works, quantum cosmologists appeal again to relationalist strategies to make sense of these probabilities. That is, the $W(a,\phi;a'\phi')$ are interpreted as the probability of finding a universe of size $a$ when the field takes the value $\phi$ given that it had size $a'$ when the field was $\phi'$. Clearly, this suffers from the same conceptual shortcomings as the relational strategy discussed in the previous section, and therefore it can be argued to be based on the conceptual mistake of taking dynamical variables to be time variables. 

In this sense, I find that probabilistic approaches either fall back on the relational strategy and its shortcomings or leave it unclear the way they are to be interpreted and connected with the classical theory we started with. In either case, the probabilistic approach does not seem to offer a compelling solution for the problem of time for quantum cosmology.

\subsection{Semiclassical approaches?}

Finally, there is a popular approach in the literature that focuses just on the discussion of a class of states, semiclassical states.\footnote{See \citep{Kiefer2012} and the philosophical discussions in \citep{Chua2021} and \citep{huggett_finding_2023}.} In this sense, this approach is limited and it does not offer an interpretation of full quantum cosmology. For this reason, it is not an approach that is very helpful in solving the problem of time of quantum gravity in general and it does not provide an illuminating way forward to interpret timeless states in general. On the other hand, of the approaches discussed in this article is the one that gets closer to recovering the temporal structure of the original model, even if it is only because it is somewhat artificially imposed. 

In semiclassical approaches one looks at only a subset of all the possible solutions to the constraint equation. These are states that satisfy a number of technical conditions, which can roughly speaking be expressed as the requirement that states are peaked along a one-dimensional curve in configuration space, at least for some regions. These states are directly interpreted as describing a classical universe evolving along this one-dimensional trajectory, $a(\tau),\phi(\tau)$. One therefore `reads' the chrono-ordinal structure from the shape of the state in configuration space. 

One example of this kind of state was originally discussed in \citep{kiefer_wave_1988,Kiefer2012}. This is the canonical quantization of a spatially closed cosmology with a massless scalar field. One can show that there exist solutions to the constraint equation of this model which are peaked along the trajectories that the universes described by the classical model follow. In this sense, it is tempting to interpret the quantum state to describe the same classical universe as the classical trajectory, including the same temporal structures.

By following the peak of the wavefunction in configuration space one gets a one-dimensional curve, and one can take any parametrization of this curve to give the chrono-ordinal structure of time. Chrono-metricity is slightly more complicated to obtain. In some discussions like the one in \citep{huggett_finding_2023} one just postulates that chrono-metricity is recovered by identifying one of the configuration variables as time and defining the chrono-metric aspect of time as the variation of that variable. In their this discussion, they take $a$\footnote{Actually they take, $\alpha$, a function of $a$ to be time, but nothing essentially changes by taking $a$.} to be time, and they claim that the variation of $a$ defines a metric for time. However, from my discussion above it should be clear that variation of $a$ may correlate with time, but it is not time and it does not give back the prediction of the age of the universe. Saying that the universe has expanded $x$ times doesn't give us the information about how long it has taken, i.e., it does not tell us if this has happened in 13.8 billion years or not.

The semiclassical aproach can do better. One can show that the peak of semiclassical states approximately satisfies a Hamilton-Jacobi equation.\footnote{See the discussion in \citep{Kiefer2012} and \citep{Chua2021}.} This equation is equivalent to a semiclassical version\footnote{By this I mean that this is a quantum-corrected version of Einstein equations known as semiclassical gravity.} of Einstein equations and it allows recovering a lapse function\footnote{In the case of full general relativity it is argued that a shift vector can also be recovered.} and hence recovering the metric aspect of time. In other words, for semiclassical states one can claim that as they satisfy the same (or close enough) equations as classical trajectories, then time will be just the same. If the peak of a semiclassical state follows the trajectory of a universe that expands in 13.8 billion years, then one claims that it describes that universe evolving in the same time.

In this sense, in the semiclassical interpretation of quantum cosmology, one takes a class of states and postulates a temporal structure for them. In this way, the problem of time is claimed to be solved by this postulation. However, there are several worries with this approach.

First, this approach says nothing about how to interpret the states or regions in which the semiclassical approximations. When proponents of the semiclassical approach discuss general states, they tend to fall into either the relational or probabilistic interpretations, which I have argued above are not satisfactory. In this sense, as an interpretation is at best incomplete. Importantly, in some states there are regions in which the approximation holds which end in regions in which states are no longer peaked. One could argue that in these regions is where interesting quantum cosmological phenomena happen, but the semiclassical interpretation stops working at that point, and hence our theory would stop being useful precisely in the situations in which we are most interested in understanding.

Second, the worry with the choice of clock in the relational interpretation also applies to the semiclassical interpretation. The states discussed in \citep{Gielen2022,Gielen2022a} are semiclassical solutions for the same model, and hence we also have the problem that there generally is not a unique possible semiclassical state. In this sense, the semiclassical approach gives different predictions depending on which states one takes to be relevant.

Third, one may also be in the opposite situation, i.e., that there are no semiclassical states or regions to start with. This seems to be the case with our toy model with two particles at least for certain potentials. In this case we find states which typically will not have the shape of a semiclassical state. For instance, for quadratic potentials we could have that the only solution to the constraint equation is the ground state of the harmonic oscillator for both particles. This is just a peaked function around the equilibrium positions of both oscillators. In this sense, it is not a function peaked along a one-dimensional curve but around one point. Similarly, excited states will also contain some point-like peaks, but we will not find the one-dimensional structure that we need for the semiclassical interpretation to get started. 

We have seen that for some models the semiclassical interpretation cannot work as there are no states with the required formal properties. In the case of more general or realistic cosmological models the same may happen, and hence the semiclassical interpretation may not be applicable. For this reason, even if the semiclassical interpretation may capture some interesting physics, it is not an interpretation that solves the problem of time in full generality and maybe not even in the cosmological case.

For these reasons, while the semiclassical interpretation offers a temporal structure more similar to the one of the classical theory, its limited range of applicability, its ultimate failure to interpret general states, and the fact that this temporal structure is just postulated in an ad-hoc manner make it the case that it does not represent a satisfactory solution for the problem of time that allows claiming that quantum cosmological models represent successful theories that we understand.


\section{Conclusions}

The study of the problem of time in the cosmological setting has allowed for a simpler discussion as we have been able to leave the technical details at a side and focus on the conceptual aspects of the problem. In this sense, I have argued that the temporal structures of classical cosmological models are absent from the result of their canonical quantization, and that this is problematic for the three main interpretations of quantum cosmological models.

My analysis here agrees with the analysis in \citep{MozotaFrauca2023}: while relationalist interpretations or probabilistic interpretations may work to solve the problem of time for deparametrizable models like the second model for the toy model of two particles, they fail to do so for non-deparametrizable ones like the third model, general relativity, or cosmological models. The reason for this is that either some dynamical variables are wrongly attributed temporal features or that we are left with just some unclear interpretations that cannot be easily connected with the classical model.

Time in the classical model defined, at least, a chrono-ordinal and chrono-metric structure. The chrono-metric structure is missing or mistaken in both the relational and probabilistic interpretations, and can only be postulated for a limited set of models and states in the semiclassical approach. For this reason, I think that we lack a coherent interpretation of quantum cosmological models which allows recovering temporal structures in a consistent manner. Just as in the case of other non-deparametrizable reparametrization invariant models, the problem of time raises the worry that canonical quantization methods may not be the right way of finding a quantum theory that has classical cosmology as some sort of classical limit.

\bibliographystyle{agsm}
\bibliography{quantum_cosmology.bib}

@article{Gryb2016,
	title = {Time remains},
	volume = {67},
	issn = {14643537},
	doi = {10.1093/bjps/axv009},
	number = {3},
	journal = {British Journal for the Philosophy of Science},
	author = {Gryb, Sean and Thébault, Karim P.Y.},
	year = {2016},
	pages = {663--705},
}

@article{Ashtekar2006,
	title = {Quantum nature of the big bang: {Improved} dynamics},
	volume = {74},
	issn = {15507998},
	doi = {10.1103/PhysRevD.74.084003},
	number = {8},
	journal = {Physical Review D - Particles, Fields, Gravitation and Cosmology},
	author = {Ashtekar, Abhay and Pawlowski, Tomasz and Singh, Parampreet},
	month = oct,
	year = {2006},
	note = {arXiv: gr-qc/0607039
Publisher: American Physical Society},
	pages = {084003},
}

@book{Anderson2017,
	address = {Cham},
	title = {The {Problem} of {Time}},
	volume = {190},
	isbn = {978-3-319-58846-9},
	publisher = {Springer International Publishing},
	author = {Anderson, Edward},
	year = {2017},
	doi = {10.1007/978-3-319-58848-3},
	note = {Series Title: Fundamental Theories of Physics
Publication Title: Fundamental Theories of Physics
ISSN: 23656425},
}

@article{Ashtekar2009,
	title = {Loop quantum cosmology: {An} overview},
	volume = {41},
	issn = {15729532},
	doi = {10.1007/s10714-009-0763-4},
	number = {4},
	journal = {General Relativity and Gravitation},
	author = {Ashtekar, Abhay},
	month = feb,
	year = {2009},
	note = {arXiv: 0812.0177
Publisher: Springer New York},
	pages = {707--741},
}

@incollection{Bojowald2004,
	title = {Cosmological {Applications} of {Loop} {Quantum} {Gravity}},
	publisher = {Springer, Berlin, Heidelberg},
	author = {Bojowald, Martin and Morales-Técotl, Hugo A.},
	month = jun,
	year = {2004},
	doi = {10.1007/978-3-540-40918-2_17},
	pages = {421--462},
}

@article{Rovelli2002,
	title = {Spacetime states and covariant quantum theory},
	volume = {65},
	issn = {05562821},
	doi = {10.1103/PhysRevD.65.125016},
	number = {12},
	journal = {Physical Review D - Particles, Fields, Gravitation and Cosmology},
	author = {Reisenberger, Michael P. and Rovelli, Carlo},
	month = jun,
	year = {2002},
	note = {arXiv: gr-qc/0111016
Publisher: American Physical Society},
	pages = {1250161--12501616},
}

@article{Colosi2003,
	title = {Simple background-independent {Hamiltonian} quantum model},
	volume = {68},
	issn = {15502368},
	doi = {10.1103/PhysRevD.68.104008},
	number = {10},
	journal = {Physical Review D - Particles, Fields, Gravitation and Cosmology},
	author = {Colosi, Daniele and Rovelli, Carlo},
	month = jun,
	year = {2003},
	note = {arXiv: gr-qc/0306059},
}

@article{Chua2021,
	title = {No time for time from no-time},
	volume = {88},
	issn = {1539767X},
	doi = {10.1086/714870},
	number = {5},
	journal = {Philosophy of Science},
	author = {Chua, Eugene Y.S. and Callender, Craig},
	month = may,
	year = {2021},
	pages = {1172--1184},
}

@book{Wuthrich2021,
	title = {Philosophy {Beyond} {Spacetime}},
	publisher = {Oxford University Press},
	editor = {Wüthrich, Christian and Le Bihan, Baptiste and Huggett, Nick},
	month = aug,
	year = {2021},
	doi = {10.1093/oso/9780198844143.001.0001},
	note = {Publication Title: Philosophy Beyond Spacetime},
}

@book{Calcagni2017,
	title = {Classical and {Quantum} {Cosmology}},
	isbn = {978-3-319-41125-5},
	publisher = {Springer International Publishing},
	author = {Calcagni, Gianluca},
	year = {2017},
	doi = {10.1007/978-3-319-41127-9},
	note = {Series Title: Graduate Texts in Physics},
}

@article{Kuchar1993,
	title = {Canonical {Quantum} {Gravity}},
	volume = {1992},
	journal = {General relativity and gravitation},
	author = {Kuchař, Karel V.},
	year = {1993},
	note = {Publisher: IOP Publishing Bristol, UK},
	pages = {119},
}

@inproceedings{Kuchar1992,
	address = {Singapore},
	title = {Time and interpretations of quantum gravity},
	doi = {10.1142/S0218271811019347},
	booktitle = {Proceedings of the 4th {Canadian} {Conference} on {General} {Relativity} and {Relativistic} {Astrophysics}},
	publisher = {World Scientific Publishing Company},
	author = {Kuchař, Karel V.},
	editor = {Kunstatter, G. and Vincent, D. and Williams, J.},
	month = jul,
	year = {1992},
	note = {ISSN: 02182718},
}

@book{Kiefer2012,
	address = {New York, NY},
	title = {Quantum {Gravity}},
	isbn = {978-0-19-958520-5},
	publisher = {Oxford University Press},
	author = {Kiefer, Claus},
	month = apr,
	year = {2012},
	doi = {10.1093/acprof:oso/9780199585205.001.0001},
}

@article{Isham1993,
	title = {Canonical {Quantum} {Gravity} and the {Problem} of {Time}},
	doi = {10.1007/978-94-011-1980-1_6},
	journal = {Integrable Systems, Quantum Groups, and Quantum Field Theories},
	author = {Isham, Chris J.},
	year = {1993},
	note = {arXiv: gr-qc/9210011},
	pages = {157--287},
}

@book{Huggett2020,
	title = {Beyond {Spacetime}},
	isbn = {978-1-108-65570-5},
	publisher = {Cambridge University Press},
	editor = {Huggett, Nick and Matsubara, Keizo and Wüthrich, Christian},
	month = apr,
	year = {2020},
	doi = {10.1017/9781108655705},
	note = {Publication Title: Beyond Spacetime},
}

@book{Rovelli2015,
	address = {Cambridge},
	title = {Covariant loop quantum gravity: {An} elementary introduction to quantum gravity and spinfoam theory},
	isbn = {978-1-107-70691-0},
	publisher = {Cambridge University Press},
	author = {Rovelli, Carlo and Vidotto, Francesca},
	month = jan,
	year = {2015},
	doi = {10.1017/CBO9781107706910},
	note = {Publication Title: Covariant Loop Quantum Gravity: An Elementary Introduction to Quantum Gravity and Spinfoam Theory},
}

@article{Oriti2017,
	title = {The universe as a quantum gravity condensate},
	volume = {18},
	issn = {16310705},
	doi = {10.1016/j.crhy.2017.02.003},
	number = {3-4},
	journal = {Comptes Rendus Physique},
	author = {Oriti, Daniele},
	year = {2017},
	note = {arXiv: 1612.09521},
	pages = {235--245},
}

@article{MozotaFrauca2023,
	title = {Reassessing the problem of time of quantum gravity},
	volume = {55},
	issn = {0001-7701},
	doi = {10.1007/s10714-023-03067-x},
	number = {1},
	journal = {General Relativity and Gravitation},
	author = {Mozota Frauca, \'Alvaro},
	month = jan,
	year = {2023},
	note = {arXiv: 2301.07973},
	pages = {21},
}

@article{Gielen2022a,
	title = {Unitarity, clock dependence and quantum recollapse in quantum cosmology},
	volume = {39},
	issn = {0264-9381},
	doi = {10.1088/1361-6382/AC504F},
	number = {7},
	journal = {Classical and Quantum Gravity},
	author = {Gielen, Steffen and Menéndez-Pidal, Lucía},
	month = apr,
	year = {2022},
	note = {arXiv: 2109.02660
Publisher: IOP Publishing},
	pages = {075011},
}

@article{Gielen2022,
	title = {Unitarity and quantum resolution of gravitational singularities},
	volume = {31},
	issn = {02182718},
	doi = {10.1142/S021827182241005X},
	number = {14},
	journal = {International Journal of Modern Physics D},
	author = {Gielen, Steffen and Menéndez-Pidal, Lucía},
	month = sep,
	year = {2022},
	note = {arXiv: 2205.15387
Publisher:  World Scientific Publishing Company },
}

@article{Agullo2017,
	title = {Loop quantum cosmology},
	doi = {10.1142/9789813220003_0007},
	journal = {Loop Quantum Gravity: The First 30 Years},
	author = {Agullo, Ivan and Singh, Parampreet},
	month = mar,
	year = {2017},
	note = {Publisher: World Scientific Publishing Co. Pte. Ltd.
ISBN: 9789813220003},
	pages = {183--240},
}

@unpublished{huggett_finding_2023,
	title = {Finding {Time} for {Wheeler}-{Dewitt} {Cosmology}},
	author = {Huggett, Nick and Thébault, Karim},
	year = {2023},
	note = {arXiv:2310.11072},
}

@article{kiefer_wave_1988,
	title = {Wave packets in minisuperspace},
	volume = {38},
	doi = {10.1103/PhysRevD.38.1761},
	number = {6},
	journal = {Physical Review D},
	author = {Kiefer, Claus},
	month = sep,
	year = {1988},
	note = {Publisher: American Physical Society},
	pages = {1761--1772},
}

@incollection{mozota_frauca_problem_2024,
	series = {The {SILFS} series},
	title = {The {Problem} of {Time} for {Non}-{Deparametrizable} {Models} and {Quantum} {Gravity}},
	volume = {4},
	isbn = {978-1-84890-455-2},
	booktitle = {Current {Topics} in {Logic} and the {Philosophy} of {Science}. {Papers} from {SILFS} 2022 postgraduate conference.},
	publisher = {College Publications},
	author = {Mozota Frauca, \'Alvaro},
	editor = {Bianchini, Francesco and Fano, Vincenzo and Graziani, Pierluigi},
	month = mar,
	year = {2024},
}

@incollection{mozota_frauca_time_2024,
	address = {Cham},
	title = {Time is {Order}},
	isbn = {978-3-031-61860-4},
	language = {en},
	booktitle = {Time and {Timelessness} in {Fundamental} {Physics} and {Cosmology}: {Historical}, {Philosophical}, and {Mathematical} {Perspectives}},
	publisher = {Springer Nature Switzerland},
	author = {Mozota Frauca, \'Alvaro},
	editor = {De Bianchi, Silvia and Forgione, Marco and Marongiu, Laura},
	year = {2024},
	doi = {10.1007/978-3-031-61860-4_4},
	pages = {49--67},
}

\end{document}